\documentclass[%
 aip,
 amsmath,amssymb,
 reprint,%
]{revtex4-1}

\usepackage{graphicx}% Include figure files
\usepackage{dcolumn}% Align table columns on decimal point
\usepackage{bm}% bold math
\usepackage[utf8]{inputenc}
\usepackage[T1]{fontenc}
\usepackage{mathptmx}
\usepackage{etoolbox}
\usepackage{float}

\makeatletter
\def\@email#1#2{%
 \endgroup
 \patchcmd{\titleblock@produce}
  {\frontmatter@RRAPformat}
  {\frontmatter@RRAPformat{\produce@RRAP{*#1\href{mailto:#2}{#2}}}\frontmatter@RRAPformat}
  {}{}
}%
\makeatother
\begin{document}

\preprint{AIP/123-QED}

\title[Sample title]{Mid-infrared Kerr index evaluation via cross-phase modulation with a near-infrared probe beam}
% Force line breaks with \\
\author{D. Lorenc}
 \altaffiliation[Also at ]{International Laser Centre, Ilkovicova 3, 84104 Bratislava, Slovakia}%Lines break auInternational Laser Centre, Ilkovi\v{c}ova 3, 84104 Bratislava, Slovakiatomatically or can be forced with \\
\author{Z. Alpichshev}
 
\affiliation{%
Institute of Science and Technology Austria, Am Campus 1, 3400 Klosterneuburg, Austria%\\This line break forced% with \\
}%

\date{\today}% It is always \today, today,

\begin{abstract}
We propose a simple method to measure nonlinear Kerr refractive index in mid-infrared frequency range that avoids using sophisticated infrared detectors. Our approach is based on using a near-infrared probe beam which interacts with a mid-IR beam via wavelength-non-degenerate cross-phase modulation (XPM). By carefully measuring XPM-induced spectral modifications in the probe beam and comparing the experimental data with simulation results we extract the value for the non-degenerate Kerr index. Finally, in order to obtain the value of degenerate mid-IR Kerr index we use the well-established two-band formalism of Sheik-Bahae {\it et al.}, which is shown to become particularly simple in the limit of low frequencies. The proposed technique is complementary to the conventional techniques such as z-scan and has the advantage of not requiring any mid-infrared detectors. 
\end{abstract}


\maketitle
Nonlinear optical frequency conversion has gone a long way in the past decades from the initial observation of faint second-harmonic generation in the 60’s \cite{Bloembergen1968} to the present day when nonlinear-optics-based sources of coherent broadband radiation have become a staple of a modern optics lab \cite{Cerullo2003}. Given the importance, there is a continual effort to improve the efficiency of such devices and, as one possible direction, it was demonstrated that many of the relevant nonlinear frequency-conversion phenomena, - be it high-harmonic generation \cite{Li2020, Chen2014} or the production of strong THz-range pulses through optical rectification \cite{Fedorov2020}, - become particularly efficient when the frequencies of the primary (fundamental) pumping beams happen to lie in the mid-infrared domain. Naturally, the progress within this approach is contingent upon detailed characterization of the optical properties of nonlinear materials in the infrared range.

Kerr effect (KE) is a nonlinear optical phenomenon wherein the refractive index of a material is changing as a response to the application of an external electric field. Unlike the closely related Pockels effect, KE is proportional to the square of the field and hence, does not require broken inversion symmetry in the host medium \citep{Boyd2008}. In fact, it is virtually ubiquitous and manifests itself prominently in many different contexts, being responsible for a great number of phenomena such as self- and cross-phase modulation of the beams, Kerr lensing, self-focusing, optical soliton formation, optical switching, passive mode-locking \cite{Boyd2008}, {\it etc.,} and as such has to be taken into account when designing any practical nonlinear optical application. 

The magnitude of KE is determined by the so-called nonlinear Kerr index $n_2$. On the conceptual level the problem of measuring $n_2$ is long solved, thanks to advent of techniques such as z-scan \cite{Chapple1997}, I-scan \cite{Taheri1996}, four-wave mixing \cite{Samoc1998, Pigeon:16}, nearly degenerate three wave mixing \cite{Adair1989}, two-beam coupling \cite{Kang1997} and a number of interferometric techniques ({\it e.g.} \cite{Jansonas2022}). However, in practice it may still pose considerable challenges, especially when it comes to the mid- and deep-IR wavelengths where experiments require specialized components (most notably detectors) and often are not as straightforward as they are with visible or near-IR radiation. As a result, the amount of information on mid-IR Kerr refractive indices is comparatively scarce and even in the case of some standard materials, mid-IR Kerr index has been characterized only recently for a limited number of wavelength values \cite{Jansonas2022, Ensley2019, Patwardhan2021}.

In this Letter we propose to go around this issue by dropping the challenging part of mid-IR detection altogether and use a near-IR probe to gauge the changes to optical properties of a material as a response to mid-IR radiation. To this end we study the non-degenerate Kerr effect-mediated interaction between near- and mid-IR pulses manifested as cross-phase modulation (XPM) \cite{Boyd2008} between the two. Specifically, we observe the mid-IR pump-induced spectral changes in the probe pulse from which we extract the value for non-degenerate Kerr index  $n_2(\Omega, \omega)$. We then use the well-established two band-theory of Sheik-Bahae \textit{et al} \cite{Hutchings1992} to reconstruct from  $n_2(\Omega, \omega)$ the mid-IR Kerr index. The convenience of the proposed methods is especially apparent in the limit where both pump and probe photon energies are significantly less than the band gap. In this regime the two-band model predicts that relationship between the degenerate and non-degenerate Kerr indices becomes particularly simple and universal, the only material-specific property being the band-gap width that enters as a simple scaling factor for frequencies.

\begin{figure*}
\includegraphics [scale=0.33] {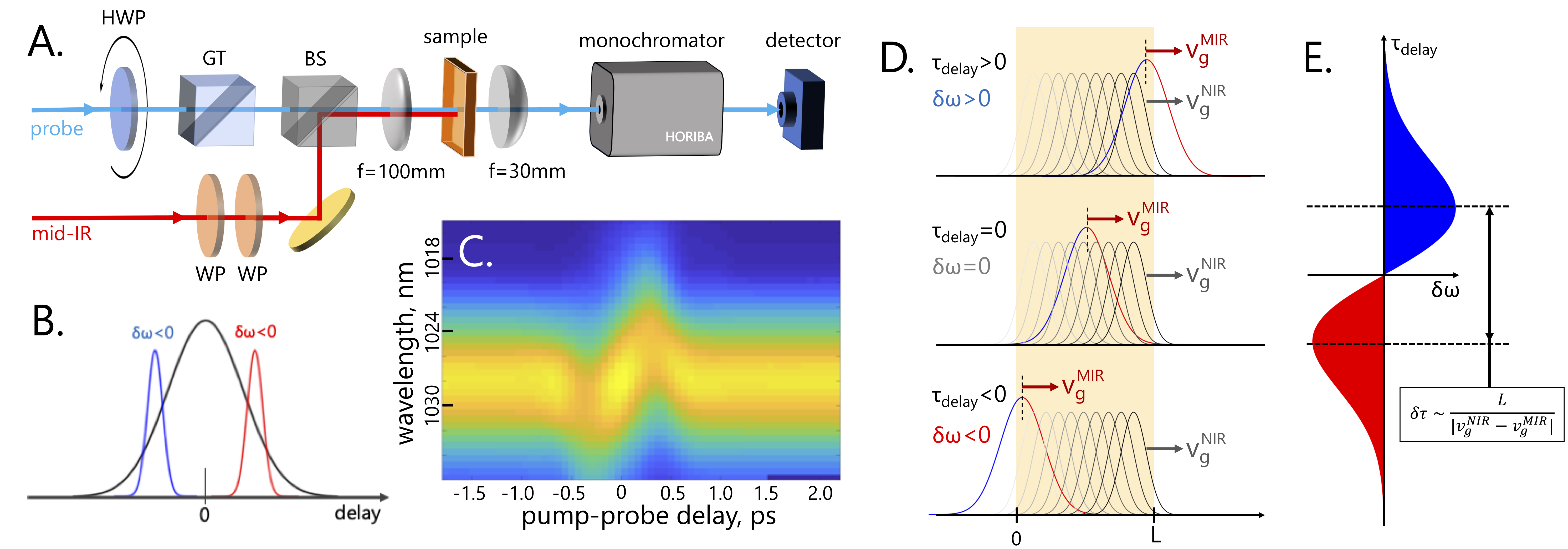}% Here is how to import EPS art
\caption{\label{fig:setup}The setup to investigate XPM-induced spectral shift. Intensity of the probe is controlled via a half-wave plate (HWP) and Glan-Taylor polarizer (GT); the intensity of the mid-IR pump coming from an OPA is tuned by a pair of wire-grid polarizers (WP). Pump and probe beams are joined by a beamsplitter (BS) and focused onto the sample with a f=100mm lens. After the sample the probe beam is re-collimated with a f=30mm lens and after passing  through monochromator detected by an avalanche photodiode detector. {\bf B)} A cartoon demonstrating cross-phase modulation (XPM) in the ideal case of no walk-off between pulses. The probe beam frequency is red- or blue-shifted depending on the sign of the Kerr index and instantaneous time derivative of pump intensity (black) (difference in pump- and probe pulse durations is exaggerated for the purpose of illustration). {\bf C)} Experimental XPM-induced spectral shift of a $\lambda_0=$1028nm probe by a $\lambda_{\mathrm{pump}}=$450 nm pump in a polycrystalline ZnSe sample as a function of pump-probe delay at peak pump intensity $I_0=2.2\times10^{14}$W/cm$^2$. {\bf D)} XPM with non-matching pump and probe pulse group velocities $v^{\mathrm{MIR}}_g$ and $v^{\mathrm{NIR}}_g$ respectively ($v^{\mathrm{MIR}}_g<v^{\mathrm{NIR}}_g$). {\bf E)} XPM-induced spectral changes for the situation laid out in panel D. Evidently, in the limit of large group velocity mismatch $\Delta v_{\mathrm{g}}$, the temporal separation between extremal shifts (blue and red) is determined by $\Delta v_{\mathrm{g}}$ rather than original pump pulse-width.}
\end{figure*}

In our study we take ZnSe as a model system due to its relatively large band gap, well-studied linear- and nonlinear optical properties \cite{Patwardhan2021,Weber2018} and low absorption in both near-IR and mid-IR spectral ranges. The sketch of the experimental setup is shown Fig.~\ref{fig:setup}A. Tunable-frequency mid-IR pulses are generated by an optical parametric amplifier (OPA; \textit{Light Conversion} Orpheus-HE) pumped by a femtosecond laser system (\textit{Light Conversion} Pharos) producing a train of pulses with a repetition rate of 3kHz; central wavelength $\lambda_0$=1028nm; pulse duration of $\tau_{\mathrm{FWHM}}$=270fs; and 2mJ/pulse. A small fraction (5\%) of the main beam from the amplifier is split off and used as a near-IR probe while the main part pumps the OPA to produce $\lambda$=4.5$\mu$m pulses used as a mid-IR pump. Pump and probe pulses are spatially and temporally overlapped inside a $d$=1.25mm-thick, polycrystalline ZnSe sample (\textit{Crystran Ltd}). After the sample, the XPM-affected probe beam is spectrally analyzed with a monochromator (\textit{Horiba} H10) connected to an avalanche photo-diode (\textit{Becker\&Hickl} APM-400-P-078). To increase the signal-to-noise ratio, the signal from the detector is first passed through a boxcar integrator (\textit{SRS} SR250) before being analyzed in a lock-in amplifier (\textit{SRS} SR830). 

\begin{figure}
\includegraphics [scale=0.47] {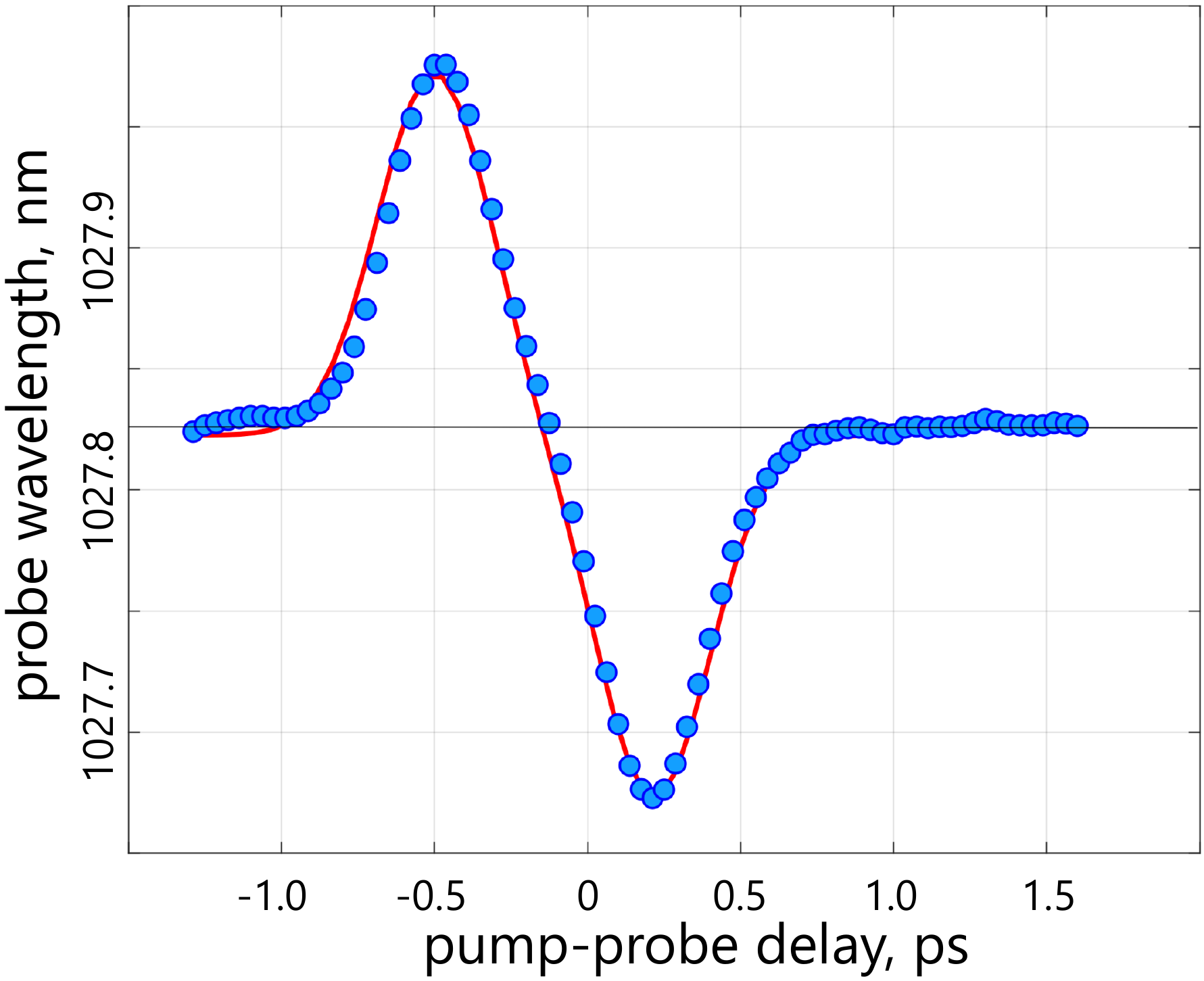}% Here is how to import EPS art
\caption{\label{fig:exp_shift} Blue dots: XPM-induced spectral shift of the central wavelength of the probe pulse as a function of pump-probe delay at peak pump intensity I$_0=2.1\times 10^{13}$W/cm$^2$; solid red curve: the result of the corresponding CGNLSE simulation.}
\end{figure}

In the experiment we record the spectrum of the probe beam at the exit from the sample for every value of pump-probe delay as shown in Fig.~\ref{fig:setup}C. In the range where pump and probe overlap temporally the spectrum of the probe is visibly perturbed by the pump. As can be seen in Fig.\ref{fig:setup}C and Fig.\ref{fig:exp_shift} the spectral shift is changing sign depending on time delay between pump and probe pulses. Qualitatively the character of the spectral change is in line with what one expects from XPM, which can be intuitively understood as shown in Fig.\ref{fig:setup}B. As a result of Kerr effect, the refractive index $n_0$ of a medium acquires a nonlinear correction $\delta n$ proportional to the intensity $I$ of the pump: $\delta n = n_2 I$, where $n_2$ is the nonlinear (Kerr) refractive index. Now, on the one hand, the total refractive index $n = n_0 + \delta n$ determines the total optical phase accumulated by the probe pulse with wavelength $\lambda = 2 \pi c/\omega$ as it propagates through the sample of thickness $L$: $\phi = 2\pi n L/\lambda$. On the other, time derivative of the phase determines the probe frequency $\omega=\dot{\phi}$. It is then clear that since time-dependent pump intensity $I(t)$ modifies $\phi(t)$, the frequency $\delta \omega$ acquires a correction proportional to the derivative: 
$$\delta \omega = 2\pi \frac{L}{\lambda} n_2 \frac{dI(t)}{dt}$$ 
\noindent Therefore analysing the value of XPM-induced spectral shift and knowing the properties of the pump pulse such as its duration and peak intensity, one can in principle expect to be able to extract the value for nonlinear Kerr index $n_2$ of the medium. 

Needless to say, this simple analysis is only valid for the artificial case of a monochromatic probe and a steadily growing pump intensity $I(t)$. In the more realistic case of pulsed pump and probe, a more involved analysis is necessary. The most important issue in a real experiment is that unlike the situation depicted in Fig.\ref{fig:setup}B, the duration of pump and probe pulses are similar, which means that different parts of the latter experience different spectral shifts. Another problem can be seen upon inspecting Fig.\ref{fig:setup}C or Fig.\ref{fig:exp_shift}. Here one can see that the points of maximum spectral shifts are separated by about $\delta \tau \approx 0.7$ps. This is considerably more than the pulsewidth of of the pump pulse ($\tau_{\mathrm{pu}}\approx 270$fs), which according to the naive cartoon in Fig.\ref{fig:setup}B should set the value for $\delta \tau$. This discrepancy comes from the fact that due to large wavelength difference, the mid-IR pump and near-IR probe pulses propagate through the sample with significantly different group velocities. Then the interaction between the pulses occurs more like the Fig.\ref{fig:setup}D ($v_g^\mathrm{NIR} > v_g^\mathrm{MIR}$). Here, in the generic situation when the two pulses meet in the middle of a sufficiently thick sample, each temporal segment of the probe (gray) experiences the entire pump (red-blue) pulse, therefore net spectral shift in the former integrates to zero. The spectral shift becomes non-zero only when the delay between the pulses is such that they overlap near one of the sample edges. The maximum spectral shifts then is achieved when the probe pulse center coincides with one of the slopes of the pump pulse. The separation between these points is consequently $\delta \tau \approx L/|v_g^{\mathrm{NIR}}-v_g^\mathrm{MIR}|$ (for $\delta \tau \gg \tau_{\mathrm{pu}}$).

To accurately take these effects into account we simulate the interaction between pulses by means of solving a system of coupled generalized nonlinear Schroedinger equations (CGNLSE) for frequency-nondegenerate fields \cite{2013} (see Supplementary Material for details). In order to simplify the calculation we ignore all absorption effects, which is justified since $\hbar \left( \omega +  \Omega \right) < \Delta_{\mathrm{gap}}$, where $\Delta_{\mathrm{gap}}$ is the band-gap of ZnSe and $\omega$ and $\Omega$ are probe and pump frequencies respectively. 

In Fig.\ref{fig:exp_shift} we plot the central wavelength of the probe as a function of delay between the pump and probe pulses (blue dots) taken with peak pump pulse intensity of $I_0=2.1\times 10^{13}$W/cm$^2$. This relatively moderate value was chosen to avoid various heating- and multi-photon-absorption-related phenomena not accounted for in our calculation. The solid red line in Fig.\ref{fig:exp_shift} is the result of a CGNLSE simulation with $n_2 \approx 0.65\times10^{-18}$\,m$^2$/W. Given the excellent agreement, we take it as the measured value for the non-degenerate nonlinear Kerr index of (polycrystalline) ZnSe.

\begin{figure}
\includegraphics [scale=0.47] {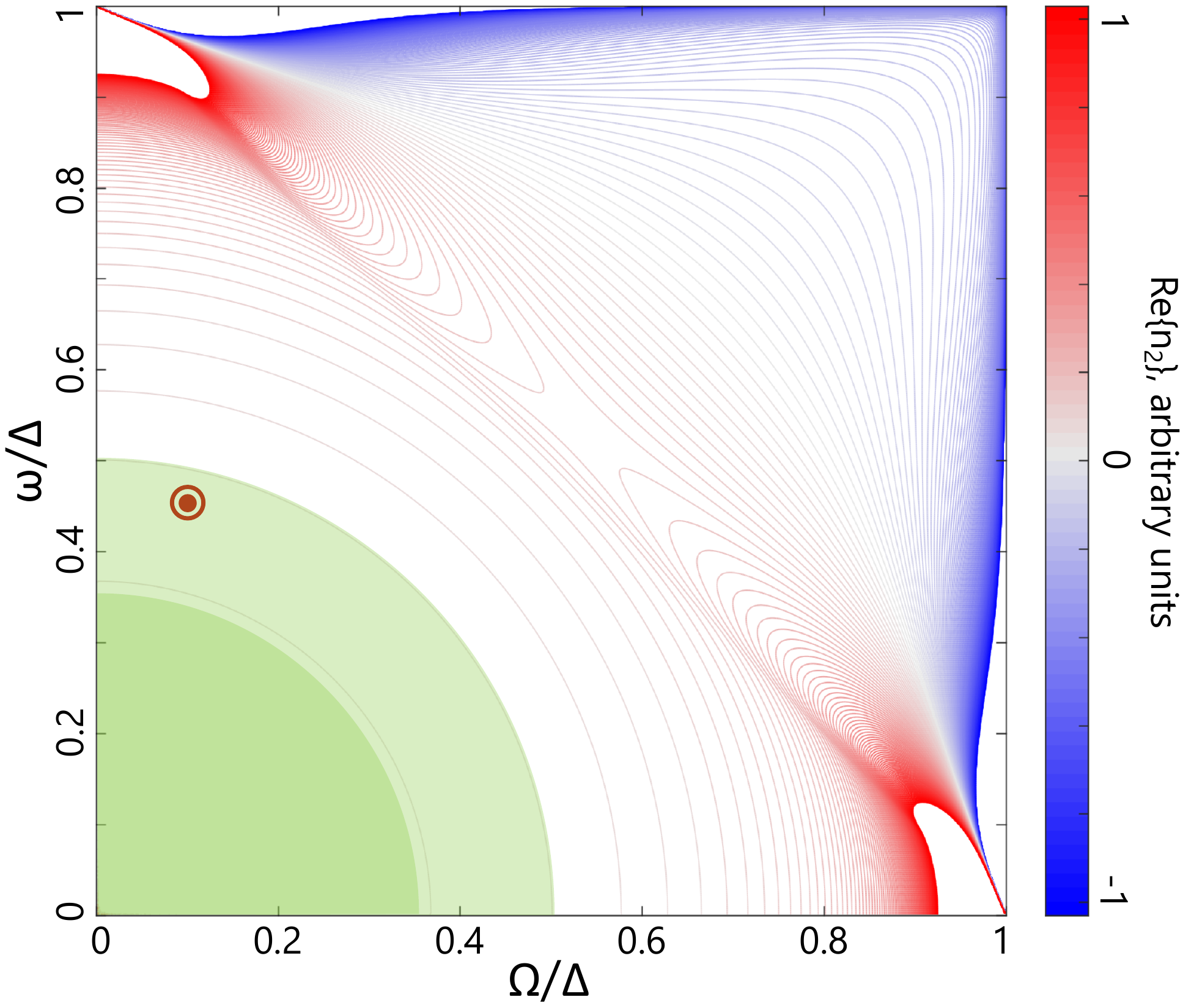}% Here is how to import EPS art
\caption{\label{fig:fig3} Real part of nondegenerate Kerr index $n_2$($\omega,\Omega$) calculated for 2-band model through dispersive analysis of Sheik-Bahae {\it et al.} In the shaded regions, the relative error of the approximate expression in eq.\ref{eq:taylor} does not exceed 2\% (darker green) and 10\% (light green); The red spot marks the point that corresponds to the conditions of the present experiment ($\omega /\Delta \approx 0.45$, $\Omega/\Delta \approx 0.10$; with ${\Delta} = 2.82$eV for ZnSe).}
\end{figure}

In the last step we need to relate the value for the non-degenerate mid-IR/near-IR Kerr index $n_2(\omega,\Omega)$ to the degenerate mid-IR value $n_2(\Omega,\Omega)$. To this end we note that generally speaking, in the limit $\omega^2,\Omega^2 \ll {\Delta^2_{\mathrm{gap}}}$ Kerr index must have the form:

\begin{equation}
n_2(\omega, \Omega) \propto  1+  \Gamma \times \left(  \frac{\omega^2 + \Omega^2 }{\Delta_{\mathrm{gap}}^2}    \right) 
\label{eq:taylor}
\end{equation}

\noindent with $\Gamma$ being some numeric factor on the order of unity whose exact value depends on the microscopic details of a given material. The form of eq.\ref{eq:taylor} is dictated by the fact that on the one hand, $n_2(\omega, \Omega)$, -- being an observable quantity, -- must be an even function of both $\omega$ and $\Omega$, and on the other, it must be symmetric against $\omega \leftrightarrow \Omega$ owing to Kleinman symmetry which holds in the low-frequency limit \cite{Boyd1999}. Having said that $\Gamma$ is not expected to differ significantly from unity, we can also calculate its value for the specific case of two-band model \cite{SheikBahae1991} which is known to adequately capture nonlinear properties of ZnSe \cite{Hutchings1992}. Fitting the expression for $n_2(\omega, \Omega)$ (see Supplementary Material) around the origin with eq.\ref{eq:taylor} gives $\Gamma = 1.35$. The accuracy of the low-frequency expansion in eq.\ref{eq:taylor} can be seen in Fig.\ref{fig:fig3} where we plot $n_2(\omega, \Omega)$, as calculated according to the two band model, and mark the experimental conditions of the present work with a red spot. As is seen, the quadratic approximation provides satisfactory agreement in a broad region around the origin which includes the wavelengths used in the present work. Using eq.\ref{eq:taylor} and the value for non-degenerate $n_2$ we obtain a value for degenerate nonlinear Kerr index of polycrystalline ZnSe $n^{\mathrm{poly}}_2\approx (0.5 \pm 0.1)\times10^{-18}$\,m$^2$/W at $\lambda$=4.5$\mu$m; the main source of error here comes from the determination of the pump beam intensity (see Supplementary Material). This number is similar in magnitude, albeit slightly less than the values obtained previously for single- \cite{Patwardhan2021} and poly-crystalline ZnSe samples \cite{Jansonas2022, Ensley2019, Werner2019}.

Lastly, one might be interested in the dependence of the magnitude of cross-phase modulation on the polarizations of pump and probe beams. The brute force approach would be to redo the spectral analysis above for all possible polarization configuration. Such an experiment is certainly doable, albeit time consuming. However, as we show below, it is in fact not necessary when one is not after absolute values for susceptibilities, but is only interested in the ratios between the different components of the nonlinear susceptibility tensor $\chi^{(3)}_{\alpha \beta \gamma \delta}$. Since XPM was established above as the main interaction channel between pump and probe pulses, one can take it for granted. Then we note that the wavelength of the probe $\lambda=1028$nm lies in the range where the sensitivity of a Si photodiode-based detector has a strong wavelength dependence (see inset in Fig.\ref{fig:rectangle}A for the responsivity of {\it{Thorlabs}} PDA100A2 detector used here). Therefore, any pump-induced changes in the probe spectrum will be detected by it.

\begin{figure}
\includegraphics [scale=0.3] {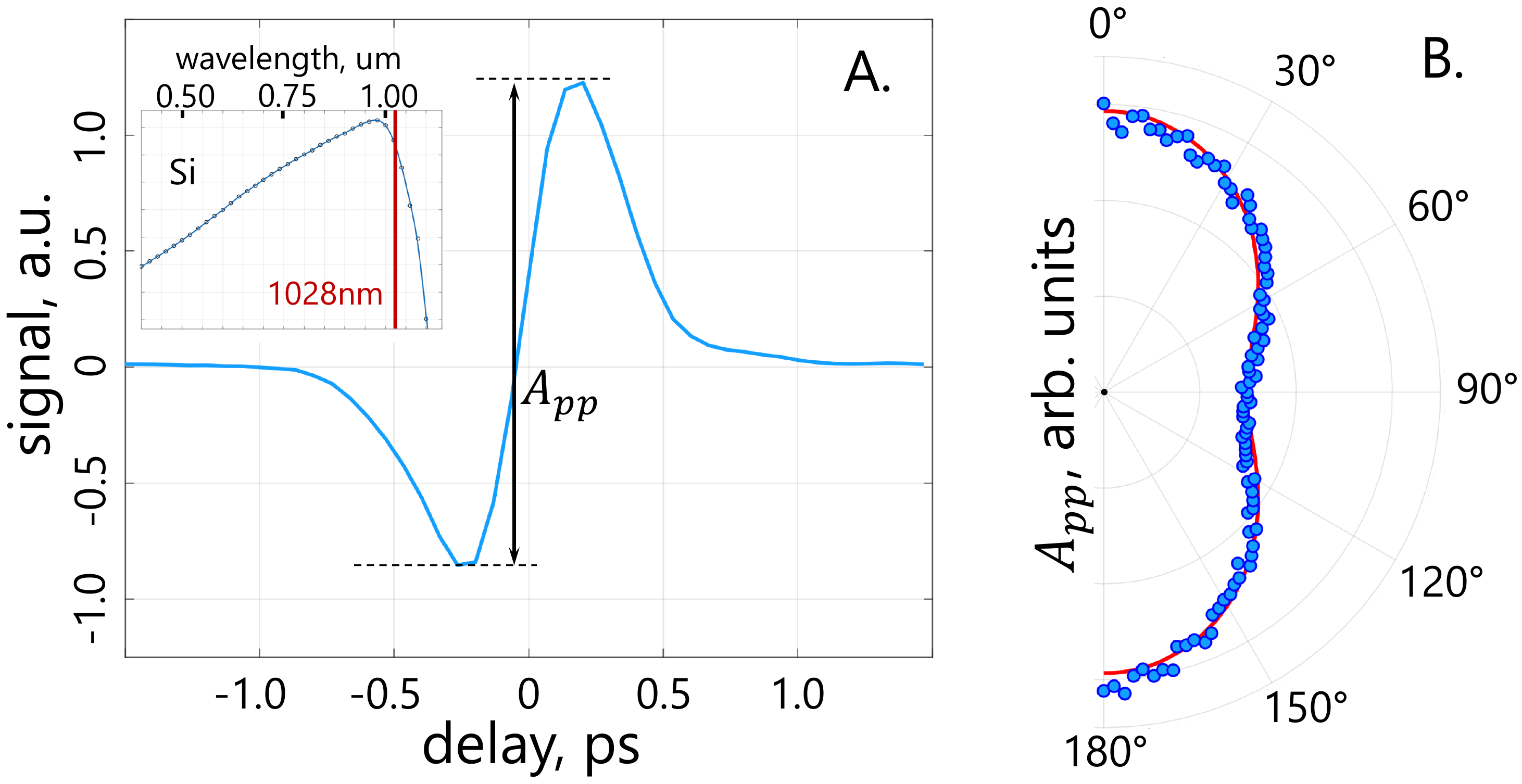}% Here is how to import EPS art
\caption{\label{fig:rectangle} Signal at the Si-diode photodetector detecting probe beam after it was modified by the pump through XPM for collinear polarization orientation of pump and probe beams; Inset: sensitivity of the Si-based photodiode used in this work as a function of wavelength \cite{Si_det}; red vertical line shows the probe beam wavelength; {\bf B)} Blue dots: peak-to-peak ($A_{\mathrm{pp}}$) amplitude of the signal at the detector as a function of angle $\theta$ between pump and probe polarizations; red line: fit to $a + b \cos(2\theta)$}
\end{figure}

In Fig.\ref{fig:rectangle}A we show the transient signal at 1028nm obtained in the standard transmission-geometry pump-probe configuration with collinear pulses having collinear polarizations (see Supplementary Material for details). As expected for XPM, the transient has a bi-polar character, qualitatively similar to Fig.\ref{fig:exp_shift}. The peak-to-peak amplitude of the signal $A_{pp}$ in Fig.\ref{fig:rectangle}A can therefore be used as a measure of XPM, and consequently Kerr index $n_2$. In Fig.\ref{fig:rectangle}B we plot this quantity as a function of the angle $\theta$ between pump and probe polarizations (blue dots). This simple angular dependence of the Kerr index $n_2(\theta)$ in a polycrystalline sample is natural and can be used to find relations between the various components of the susceptibility tensor. Indeed, when expressed in terms of crystalline nonlinear susceptibility $\chi_{\alpha \beta \gamma \delta}$, the effective nonlinear susceptibility of a polycrystalline sample $\chi_{\mathrm{eff}}(\theta)$ must have the following form (see Supplementary Material):

\begin{multline}
n_2(\theta) \propto \chi_{\mathrm{eff}}(\theta) = \frac{1}{2}\left(\chi_{xxxx}+\chi_{xxyy}\right) +\\ +\frac{\cos(2\theta)}{4} \left(  \chi_{xxxx}-\chi_{xxyy} +\chi_{xyyx}+\chi_{xyxy} \right)
\label{eq:fitting}
\end{multline}

The red solid curve in Fig.\ref{fig:rectangle}B is a fit $\chi_{\mathrm{eff}}(\theta) = a+b \cos(2\theta)$ producing an experimental value for $b/a \approx 1/3$ that can be used to obtain the ratio between the quantities in the parentheses. For this we note that for a cubic crystal structure of ZnSe $\chi_{xxxx} \approx \chi_{xxyy}+\chi_{xyyx}+\chi_{xyxy}$ \cite{Burns1971}, and find that for ZnSe at 4.5$\mu$m pump and 1.03$\mu$m probe, $\chi_{xxyy} \approx \chi_{xxxx}/2$ and $\chi_{xyxy}=\chi_{xyyx} \approx \chi_{xxxx}/4$.

In conclusion, we introduce an alternative technique to measure the nonlinear Kerr index in the mid-infrared range by studying the effects of cross-phase modulation on a secondary near-infrared probe beam. Employing this near-IR beam as a probe allows circumventing the necessity for sophisticated IR detectors necessary for conventional methods such as z-scan. In order to relate the measured non-degenerate and the sought-after degenerate mid-IR Kerr indices, we analyze the frequency dependence of nonlinear refractive index and with the help of nonlinear dispersive analysis of Sheik-Bahae {\it et al.}, establish a general expression for the frequency dependence of non-degenerate $n_2(\omega, \Omega)$ that is not limited to a specific material. In the proof-of-principle experiment we measured the Kerr index for a polycrystalline ZnSe sample to find $n_2 \approx (0.5\pm 0.1)\times10^{-18}$\,m$^2$/W at 4.5$\mu$m.
\\
See the supplementary material for supporting content.

\begin{acknowledgments}
The work was supported by IST Austria. The authors would like to gratefully acknowledge the help and assistance of Prof. John M. Dudley.
\end{acknowledgments}

\section*{Author Declarations}
\noindent\textbf{Conflict of Interest}
\\
The authors have no conflicts to disclose.
\\
\textbf{Author Contributions}
\\
\textbf{D. Lorenc:} Investigation (equal); Methodology (equal); Visualization (equal); Writing (equal). \textbf{Z. Alpichshev:}  Investigation (equal); Methodology (equal); Visualization (equal); Writing (equal).

\section*{Data Availability Statement}

The data that support the findings of this study are available from the corresponding author upon reasonable request.

\section*{References}
\bibliography{refs_ZnSe}% Produces the bibliography via BibTeX.

\cleardoublepage

\noindent \textbf{CGNLSE modeling.} The XPM was modeled using the usual set of coupled nonlinear Schroedinger equations  \cite{2013}:

\begin{equation}
\frac{\partial A_{1}}{\partial z}+\frac{1}{v_{g1}}\frac{\partial A_{1}}{\partial z}+i\frac{\beta_{21} }{2}\frac{\partial^2 A_{1}}{\partial z^2}= i\gamma_{1}(\left | A_{1} \right |^{2}+2\left | A_{2} \right |^{2})A_{1}
\label{eq:CGNLSE1}
\end{equation}        

\begin{equation}
\frac{\partial A_{2}}{\partial z}+\frac{1}{v_{g2}}\frac{\partial A_{2}}{\partial z}+i\frac{\beta_{22} }{2}\frac{\partial^2 A_{2}}{\partial z^2}= i\gamma_{2}(\left | A_{2} \right |^{2}+2\left | A_{1} \right |^{2})A_{2} 
\label{eq:CGNLSE2}
\end{equation}

\noindent with $A_{x}$ being the field amplitudes, $v_{gx}$ corresponding group velocities and $\beta_{2x}$ and $\gamma_{x}$ being GVD coefficients and nonlinear parameters respectively. Here we ignore all absorptive effects, justification being that on the one hand the sample is relatively thin to ignore linear absorption, and that $\Omega_{\mathrm{pump}}+\omega_{\mathrm{probe}}<\Delta_{\mathrm{gap}}$ so we can also ignore two-photon absorption. The set was integrated by using the split-step Fourrier method.

In the course of analyzing the beam interaction for various parameter values, we have also observed that for the conditions of the present experiment, group velocity dispersion (GVD) has no observable effect on the simulation outcome when corresponding coefficients in CGNLSE equations above were taken at their typical values reported in literature (as in e.g. J. Connolly et al., {\it Proc. SPIE} {\bf 181}, 141 (1979)). Therefore it is even possible to simply put GVD coefficients $\beta_{2i}$ equal to zero under these conditions. 
\vspace{5mm}

\noindent \textbf{Full expression for $n_2(\omega, \Omega)$.} The approach closely follows the treatment by Sheik-Bahae et.al.\cite{Hutchings1992}. The two-band model expression for absorption (with Raman terms included) is used to recover the real Kerr index $n_2$ by means of extended Kramers-Kronig analysis \cite{SheikBahae1991}. For reference we write down here the full expression for $n_2(\Omega, \omega)$ calculated in this approximation (two-band model, two-photon absortion and Raman):

\begin{gather}
n^{\mathrm{2PA+Raman}}_2(\omega, \Omega) \propto \frac{\Delta^{9/2}}{\omega^4\Omega^4} \cdot \left\{ {\left(\Delta -\omega - \Omega   \right) ^{3/2}} \left( {\omega} +{\Omega} \right)^2  + \right. \nonumber \\
\left. +\left( \Delta + \omega - \Omega  \right) ^{3/2} \left( {\Omega}- {\omega} \right)^2 + \right.\nonumber \\
\left. +\left(\Delta  - \omega  + \Omega  \right) ^{3/2} \left( {\omega} -{\Omega} \right)^2  + \right. \nonumber \\
\left. +\left( \Delta + \Omega + \omega   \right) ^{3/2} \left( -{\Omega}- {\omega} \right)^2 - \right. \nonumber \\
\left. - 2 \Omega^2 (\Delta-\Omega)^{3/2} \left[   1+ \frac{\omega^2}{\Omega^2} - \frac{3\omega^2}{\Omega (\Delta -\Omega)} +\frac{3}{8} \frac{\omega^2}{(\Delta -\Omega)^2}  \right]- \right. \nonumber \\
\left. - 2 \Omega^2 (\Delta+\Omega)^{3/2} \left[   1+ \frac{\omega^2}{\Omega^2} + \frac{3\omega^2}{\Omega (\Delta +\Omega)} +\frac{3}{8} \frac{\omega^2}{(\Delta +\Omega)^2}  \right]- \right. \nonumber \\
\left. -2\omega^2 (\Delta - \omega)^{3/2} \left[  1+ \frac{\Omega^2}{\omega^2} - \frac{3 \Omega^2}{\omega (\Delta -\omega)} +\frac{3}{8} \frac{\Omega^2}{(\Delta - \omega)^2}  \right]- \right. \nonumber \\
\left. -2\omega^2 (\Delta + \omega)^{3/2} \left[  1+ \frac{\Omega^2}{\omega^2} + \frac{3 \Omega^2}{\omega (\Delta +\omega)} +\frac{3}{8} \frac{\Omega^2}{(\Delta + \omega)^2}  \right]+ \right. \nonumber \\
\left.+4\Delta^{3/2} \left( \omega^2 + \Omega^2 \right) + 9 \frac{\omega^2 \Omega^2}{\Delta^{1/2}} \right\}
\label{eq:full}
\end{gather} 

where both $\omega$ and $\Omega$ are complex variables. In the degenerate limit $\omega$ = $\Omega$ the above expression can be checked to coincide with the functional dependence of $n_2(\omega, \Omega)$ given in \cite{Hutchings1992}.

\vspace{5mm}
\noindent \textbf{Pump-probe setup.} The beam delivery side remained the same as in the spectrally-resolved experiment, however, unlike the previous case we chop the 4.5$\mu$m pump beam. The probe beam interacts with pump throughout the sample and is sampled by a Si-diode-based photodetector ({\it{Thorlabs}} PDA100A2). To increase the signal-to-noise ratio, the signal is pre-processed through a boxcar integrator (\textit{SRS} SR250) before being analyzed as usual by a lock-in amplifier (\textit{SRS} SR830) synchronized with the optical chopper modulating the pump intensity.

\vspace{5mm}

\noindent \textbf{Polarization dependence of third-order susceptibility in polycrystalline ZnSe.} In order to derive an expression for the polarization dependence of $\chi^{(3)}$ and consequently $n_2$ in a polycrystalline sample, we will first write down the expression for the third-order susceptibility $\chi^{(3)}_{\alpha \alpha \beta \beta}(\omega; \omega, \Omega, -\Omega)$ in a single crystal sample for arbitrary polarization orientations of the pump and probe. Here $\omega$ ($\Omega$) is the frequency of the probe (pump) beams while $\alpha$ ($\beta$) mark the orientation of the polarization of the probe (pump) beams with respect to the optical axis of the crystal. By using the standard transformation properties of the susceptibility tensor, it is straightforward to write:

\begin{gather*}
\chi^{(3)}_{\alpha \alpha \beta \beta}= \left[\cos^2(\alpha)\cos^2(\beta) +  \sin^2(\alpha)\sin^2(\beta)\right]\cdot \chi_{xxxx}\\
+\left[\cos^2(\alpha)\sin^2(\beta) +  \sin^2(\alpha)\cos^2(\beta)\right]\cdot \chi_{xxyy} +\\
+2\sin(\alpha)\cos(\alpha)\sin(\beta)\cos(\beta) \cdot \left[ \chi_{xyxy}+\chi_{xyyx} \right]+\\
+\left[ \chi_{xxxy}+\chi_{xxyx}\right]\cdot \sin(\beta)\cos(\beta) + \left[ \chi_{xyxx}+\chi_{yxxx}\right]\cdot \sin(\alpha)\cos(\alpha)
\end{gather*} 

\noindent here we used the fact that the material in question (ZnSe) has cubic symmetry, therefore $\chi_{xxxx}=\chi_{yyyy}$, $\chi_{xyxy}=\chi_{yxyx}$, {\it etc}. Now to find the susceptibility $\chi_{\mathrm{eff}}(\theta)$ of a polycrystalline sample, where $\theta = \beta-\alpha$, we assume that the single-crystal domains are oriented with a random distribution across the sample, therefore

\begin{gather*}
\chi_{\mathrm{eff}}(\theta) = \int \frac{d\alpha}{2 \pi} \, \,\chi^{(3)}_{\alpha \alpha (\alpha+\theta)(\alpha+\theta)}=\\
=\frac{1}{2}\left(\chi_{xxxx}+\chi_{xxyy}\right) +\frac{\cos(2\theta)}{4} \left(  \chi_{xxxx}-\chi_{xxyy} +\chi_{xyyx}+\chi_{xyxy} \right)
\end{gather*} 

\noindent reproducing eq.2 of the main text. 
 
\vspace{5mm}
\noindent \textbf{Intensity measurements}. The actual intensities as given in the text and used in the CGNLSE simulation were calculated by measuring: 

1.) the respective beam waist diameters using the knife-edge technique:

%%%%%%%%%%%%%%%%%%%%%%%%%%%%%%%
\begin{figure}
\includegraphics[scale=0.4]{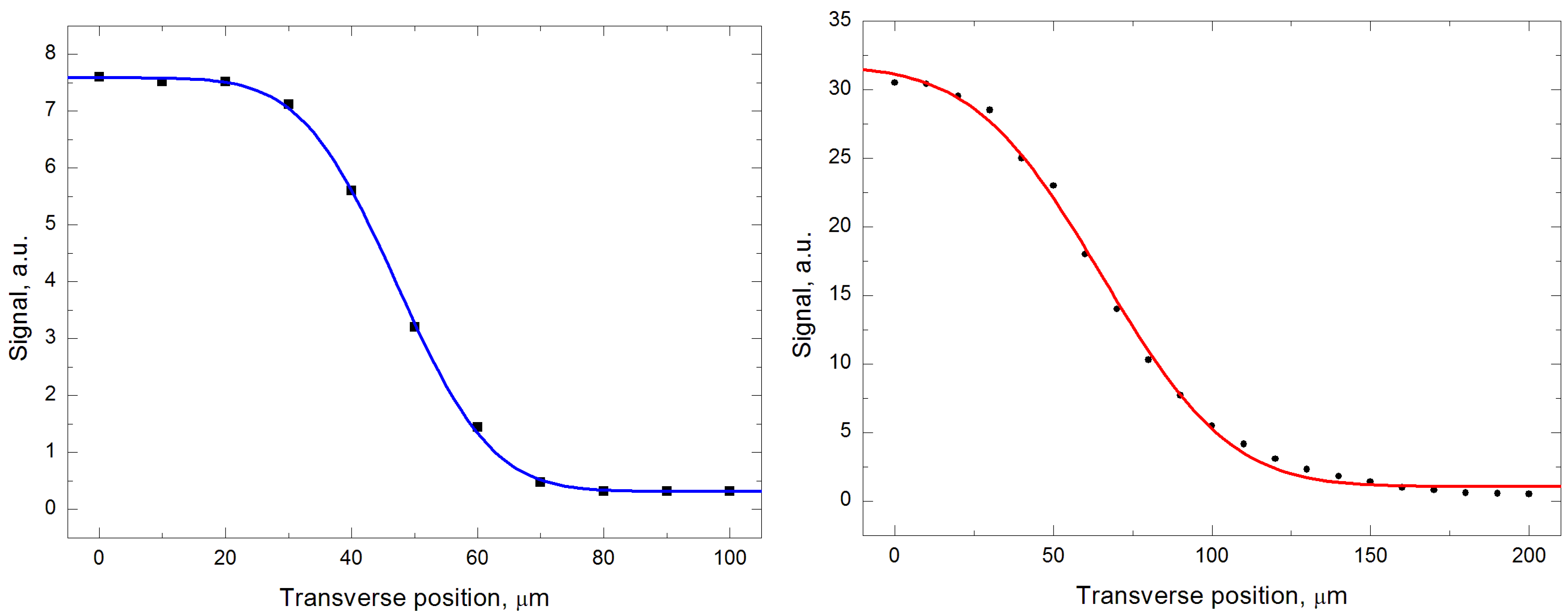}
\caption{Transverse scan of the 1030 nm probe beam (left) and the 4500 nm pump beam}
\label{fig:rectangle}
\end{figure}
%%%%%%%%%%%%%%%%%%%%%%%%%%%%%%%

\noindent with the corresponding beam-waist diameters of 24$\pm$ 1.4$\mu$m and  63$\pm$ 6.1$\mu$m for the 1030 nm probe beam and 4500 nm pump beam respectively. Note that the knife-edge scan as performed is a 1D technique and hence we estimate an additional $\pm$10$\%$ may have arisen as a result of the pump beam ellipticity.   
\\

2.) the actual power. Beam power was measured by means of an OPHIR VEGA powermeter with a 3A-P-V1 thermal head and an estimated uncertainty of $\pm$10$\%$. Hence we estimate the total uncertainty of the intensity measurements on the level of $\pm$30$\%$

\end{document}